%% file: Splitting-arXiv.tex
\documentclass[12pt]{article}                     
\usepackage{graphicx}\usepackage{amsmath,amssymb}
\usepackage{color}
\usepackage{greekbf}
\newcommand\bbeta{\mathbf{\beta}}

\begin{document}


\begin{center}

\end{center}
\input{caratteri}

\newcommand\email[1]{\texttt{#1}}
\newcommand\at{:}

\begin{center}
 {\bf \Large
Energy Splitting Theorems for Materials with Memory}
\end{center}
\medskip

\begin{center}
{\large Antonino Favata \quad
        Paolo Podio-Guidugli \quad
        Giuseppe Tomassetti
}\end{center}

\begin{center}
 \noindent Dipartimento di Ingegneria Civile, Universit\`a di Roma Tor Vergata\footnote{Via Politecnico 1, 00133 Rome, Italy. \\
{\null} \quad \ Email:
\begin{minipage}[t]{30em}
\email{favata@ing.uniroma2.it} (A. Favata)\\
\email{ppg@uniroma2.it} (P. Podio-Guidugli)\\
\email{tomassetti@ing.uniroma2.it} (G. Tomassetti)
\end{minipage}}
\small
\end{center}
\medskip

\begin{abstract}
\noindent We extend to materials with fading memory and materials
with internal variables a result previously established by one of
us for materials with instantaneous memory: the additive
decomposability of the total energy into an internal and a kinetic
part, and a representation of the latter and the inertial forces
in terms of one and the same mass tensor.
\medskip

\end{abstract}

\section{Introduction}
\label{intro} The purpose of this paper is to extend to two
classes of materials with memory a result established in
\cite{PPG} for materials that, as exemplified by standard
thermoelastic materials, can only respond
 to the current values of their state variables.

The result we aim to extend is called in \cite{PPG} the Energy
Splitting Theorem: it is shown that the (total) energy and the
inertia force have consistent representations, under the
assumptions that (i) the power expenditure of the inertia force be
linear in the velocity; and that (ii) the inertial power plus the
rate of change of the energy be translationally invariant.
More precisely, it is shown that the energy can be split in two
parts,
 \emph{internal} and \emph{kinetic}, with the internal energy independent
 of velocity and the kinetic energy a quadratic form in the velocity, based
  on a time-independent \emph{mass tensor}, the same that determines also
   the work-effective part of the inertial force.

The two material classes we here consider are: the class of
\emph{simple materials} in the sense of Truesdell and Noll
\cite{TN}, whose mechanical response is determined by the history
of the deformation gradient; and the class of materials with
\emph{internal state variables}, as considered e.g. by Coleman
 and Gurtin \cite{CG} and Lubliner \cite{Lu}, whose evolution is governed by
 a generally nonlinear differential equation (that the Energy Splitting Theorem
   had to be extendable to this material class was suggested by M.E. Gurtin in 1994,
   on reading a preprint of \cite{PPG}). Since these two material classes have a
   nonempty intersection but do not overlap, we have to prove the entry part of
   our generalized Energy Splitting Theorem twice; we give the reasons for this at
   the end of next section. Luckily, as we shall see, the rest of the proof is not
   as sensitive to the chosen class.

Our paper is organized as follows. In Section 1, we introduce the
quantities that are object of a constitutive prescription, we
stipulate their invariance properties, and we summarize the Energy
Splitting Theorem which we aim to generalize. In Section 2, we
provide a constructive proof of the Energy Splitting Theorem under
the assumption that the constitutive functionals be smooth
relative to a norm having
 the fading-memory property. In Section 3, we sketch a proof of the Energy
 Splitting Theorem for materials with internal variables.  Apart for some
  technicalities that we   try and explain carefully when they come about, the
   structure of the proofs we give is the same as the variant of the proof in \cite{PPG} given in \cite{PPG2}.

\section{Setting the stage}
\label{sec:1}

We work in a referential setting. To begin with, we introduce two
scalar volume densities, of the (total) \emph{energy}, denoted by
$\tau$, and of the \emph{inertial power}:
\[
\pi^{\rm in}=\db^{\rm in}\cdot \vb,
\]
where $\vb$ is the velocity vector and $\db^{\rm in}$ is the
\emph{inertia force} vector. Next, we define the \emph{internal
power} density to be
\begin{equation}\label{alfa}
\alpha=\dot\tau+\pi^{\rm in}
\end{equation}
(a superposed dot denotes time differentiation). Both $\tau$ and
$\db^{\rm in}$ are  constitutively prescribed at a later stage. As
now, it suffices for us to stipulate that, in principle, they both
depend on one and the same list of \emph{state variables}, that we
split as follows: $(\Lambda,\vb)$, where the list $\Lambda$
includes only \emph{translationally invariant} variables.
Precisely, a \emph{translational change in observer} is a mapping
leaving the time line unchanged:
\[
(t,x)\mapsto (t,x)^+=(t,x^+),
\]
such that, at some fixed time $\bar t$, the current shape of the
body under study is pointwise preserved, while the velocity field
varies by a uniform amount:
\begin{equation}\label{coo}
\begin{aligned}
x\mapsto x^+&=x+(t-\bar t)\wb,\\
\vb\mapsto \vb^+ &=\vb+\wb,
\end{aligned}
\end{equation}
for some fixed vector $\wb$. Thus, as to state-variable pairs,
\[
(\Lambda,\vb)\mapsto (\Lambda,\vb)^+=(\Lambda,\vb+\wb).
\]

This is the result of \cite{PPG} we generalize in the next two
sections.
\medskip

\noindent\textbf{Energy Splitting Theorem.} {\em  Let the inertial
power be linear in the velocity, in the sense that there is a
mapping $\;\Lambda\mapsto \widehat\db_0(\Lambda)\,$, referred to
as the \emph{work-effective inertia force mapping}, such that
\begin{equation}\label{inert}
\widehat\db^{\rm
in}(\Lambda,\vb)\cdot\vb=\widehat\db_0(\Lambda)\cdot\vb,\quad
\forall \;(\Lambda,\vb).
\end{equation}
Moreover, let the internal power be invariant under translational
changes in observer:
\begin{equation}\label{ip}
\hat\alpha(\Lambda^+,\vb^+)=\alpha^+=\alpha=\hat\alpha(\Lambda,\vb).
\end{equation}
Finally, let the constitutive functions $\widehat{\mathbf d}^{\rm
in}$ and $\hat\tau$ be, respectively, continuous and
twice-continuously differentiable. Then, the energy $\tau$ and the
inertial force $\db^{\rm in}$ have consistent representations,
parameterized by

 \noindent (i)  the
{\em mass tensor}, a symmetric tensor $\Mb$, independent of
$(\Lambda,\vb)$ and obeying the {\it mass conservation law}
\[
    \dot\Mb=\mathbf{0}\,;
\]
(ii) the {\em internal energy}, a scalar-valued mapping
$\Lambda\mapsto \hat\epsilon(\Lambda)$ defined over the state
space.\hfill\break \noindent These representations are:
\begin{equation}\label{reppre}
    \hat\tau(\Lambda,\vb)=\tilde\epsilon(\Lambda)+{1\over 2}\,\vb\cdot \Mb\vb\,,
\end{equation}
\begin{equation}\label{finrep}
    \tilde\db^{\rm in}(\Lambda,\vb)=-\Mb\dot \vb + \widetilde\Db^{\rm in}(\Lambda,\vb)\vb,
\end{equation}}
with $\widetilde\Db^{\rm in}(\Lambda,\vb)$ a skew-symmetric
tensor.\vskip 4pt

Note that \eqref{alfa}, \eqref{inert}, and \eqref{coo}, imply that
the invariance requirement \eqref{ip} takes the form:
\begin{equation}\label{timediff}
\frac{d\,}{dt}\Big(\hat\tau\big(\widehat\Lambda(t),\hat\vb(t)+\wb\big)-\hat\tau\big(\widehat\Lambda(t),\hat\vb(t)\big)\Big)+\widehat\db_0\big(\widehat\Lambda(t)\big)\cdot\wb=0,
\end{equation}
for every \emph{constitutive process} $t\mapsto
(\widehat\Lambda(t),\hat\vb(t))$ and for every vector $\wb$. With
this in mind, we are in a position to indicate why, in the last
part of the Introduction,
 we stated that the entry part of the proof of a theorem of this sort depends on the material
 class for which it is meant to hold:
the first and crucial step in the proof is to achieve a
preliminary additive splitting of the energy into an internal
part, that does not depend on velocity, and a kinetic part. To
take that step, it is necessary to compute the derivative of
$\hat\tau$ with respect to its first argument. This is easy in the
case considered in \cite{PPG}. Not so when, as we here do, the
constitutive dependence of energy and work-effective inertia force
on the current value of $\widehat\Lambda$ is replaced by a
functional dependence on the \emph{history} of $\widehat\Lambda$
up to time $t$, or by the current value of
$\widetilde\Lambda=(\widehat\Lambda,\widehat\bbeta)$\footnote{at
any given body point: in this paper, we leave all space
dependencies tacit.}, with $\widehat\bbeta$ a solution of the
(generally nonlinear) ordinary differential equation  governing
the time evolution of a chosen list of \emph{internal variables}
$\bbeta$:
\[
\dot\bbeta=\fb(\Lambda,\bbeta)\,.
\]
Both replacements entail a rethinking of the structure of state
space, as to the accessibility of its points and, more
importantly, as to the possibility of giving any process an
arbitrary short continuation in time.

We will discuss these technical issues at the appropriate stage of
our developments.

\section{Materials with fading memory}
Our proof of an Energy Splitting Theorem for fading-memory
materials is constructive, and is organized in four steps.

\vskip 6pt

\emph{Step 1. Translational Invariance of the Internal Power.} We
let the space of the translationally invariant state variables be
a open set $C$ of a finite-dimensional
 inner product space $L$, and we
denote by $V$ the velocity space.
 Given a \emph{state process}  $\widehat\Lambda$ ($\equiv$ a smooth differentiable curve in
 $C$),
 its \emph{history up to time $t$}  is the mapping
\[
\Lambda^t:[0,+\infty)\rightarrow
C,\quad\Lambda^t(s):=\widehat\Lambda(t-s)\,;
\]
moreover, its \emph{past history up to time $t$} is the
restriction $\Lambda^t_{r}$ of $\Lambda^t$ to the open half-line
$(0,+\infty)$, and its \emph{instantaneous value} is
$\Lambda^t(0)=\widehat\Lambda(t)$.

On adjourning the constitutive dependence of both energy and
inertia force as appropriate to materials with fading memory, we
set:
\begin{equation}\label{constitutive}
\begin{aligned}
\tilde\tau_\vb(t)&:=\hat\tau\big(\widehat\Lambda(t),\Lambda^t_{r},\hat\vb(t)\big),\\
\widetilde\db_\vb^{\rm
in}(t)&:=\widehat\db^{\rm in}\big(\widehat\Lambda(t),\Lambda^t_{r},\hat\vb(t)\big),\\
\widetilde\db_0(t)&:=\widehat\db_0\big(\widehat\Lambda(t),\Lambda^t_{r}\big),
\end{aligned}
\end{equation}
and we consistently adjourn assumption \eqref{inert}:
\begin{equation}\label{L1}
\widetilde\db_\vb^{\rm
in}(t)\cdot\hat\vb(t)=\widetilde\db_0(t)\cdot\hat\vb(t).
\end{equation}
Consequently,  the invariance requirement \eqref{timediff} can now
be written formally as follows:
\begin{equation}\label{I1}
 \frac{d\,}{dt}\big(\tilde\tau_{\vb+\wb}(t)-\tilde\tau_\vb(t)\big)+\widetilde\db_0(t)\cdot\wb=0;
\end{equation}
for it to have a precise mathematical sense, we have to specify
the regularity of the functionals $\hat\tau$ and $\widehat\db_0$.
\medskip

\emph{Step 2. Fading-Memory Property and Chain-Rule Formula.} For
$h:(0,+\infty)\rightarrow \mathbb{R}^{+}$ a non-negative measurable
function chosen once and for all and such that
\[
 \int_0^{+\infty} |h(s)|^2\,{\rm d}s<+\infty,
\]
we denote by $L_r$ the Banach space of all measurable functions
$\Lambda_{r}:(0,+\infty)\rightarrow L$, with the norm
\begin{equation}\label{norm}
 \|\Lambda_{r}\|=\left(\int_0^\infty h(s)|\Lambda_{r}(s)|^2{\rm d}s \right)^{\frac 1
 2}.
\end{equation}

Two histories are close in the topology determined by the norm
$\|\cdot\|$ if their values are close in the recent past no matter
how far apart they are in the distant past. Thus, since the
constitutive mappings are smooth, changing the history of
$\widehat\Lambda$ in the distant past does not
 affect appreciably the instantaneous values of $\tau$ and $\db^{\rm in}$. In Coleman and Noll's
  terminology \cite{CN,TN}, introducing the norm $\|\cdot \|$ endows material class with the \emph{Fading-Memory Property}.

We let the domain of the functional $\widehat\db_0$ be an open
subset $\Cc$ of the Banach space $\Lc=L\oplus L_{r}$, and we let
the common domain of  functionals $\hat\tau$ and
$\widehat\db^{in}$ be $\Cc\oplus V$. Moreover, we require that
\emph{$\widehat{\mathbf d}^{\rm in}$ be continuous and $\hat\tau$
be twice-continuously Fr\'echet differentiable}.

As pointed out in Remark 1 of \cite{CM}, the continuous
differentiability of $\hat\tau$ and the smoothness of $t\mapsto
\widehat\Lambda(t)$ guarantee that the time derivatives in
\eqref{I1} are well defined, and that the following ``Chain-Rule
Formula'' holds true:
\[
\begin{split}
\frac{\rm d}{\rm d t}\,\tilde\tau_\vb(t)={}
&\partial_\Lambda\hat\tau(\Lambda(t),\Lambda_r^t,\vb(t))
\cdot\dot\Lambda(t)\\
{}+{}&\delta_r\hat\tau(\Lambda(t),\Lambda_r^t,\vb(t))[\dot\Lambda_r^t]+
\partial_\vb\hat\tau(\Lambda(t),\Lambda_r^t,\vb(t))\cdot\dot\vb(t).
\end{split}
\]
 Here $\partial_\Lambda\hat\tau$
and $\partial_{\mathbf v}\hat\tau$ are the partial derivatives of
$\hat\tau$ with respect to its first and third argument,
respectively, and $\delta_r\hat\tau$ is
 the unique bounded linear functional on $L_r$ satisfying
\[
\begin{split}
\hat\tau(\Lambda,\Lambda_r^t+\Phi_r,\vb)
&=\hat\tau(\Lambda,\Lambda_r^t,\vb)+\delta_r\hat\tau(\Lambda,\Lambda_r^t,\vb)[\Phi_r]+o(\|\Phi_r\|)\quad\forall
\,\Phi_r\in L_r.
\end{split}
\]
By combining the invariance requirement \eqref{I1}  with the
Chain-Rule Formula,  we obtain an expression involving both the
instantaneous value and the past history of the time derivative of
the state process $\hat\Lambda$:
\begin{equation}\label{eq:5}
\begin{split}
&\Big(\partial_\Lambda{\hat\tau}(\widehat\Lambda(t),\Lambda_r^t,
\hat\vb(t)+\wb)-\partial_\Lambda{\hat\tau}(\widehat\Lambda(t),\Lambda_r^t,\hat\vb(t))\Big)\cdot\dot{\Lambda}(t)\\
&+\Big(\delta_{r}{\hat\tau}(\widehat\Lambda,\Lambda_r^t,\hat\vb(t)+\wb)-\delta_{r}{\hat\tau}(\widehat\Lambda,\Lambda_r^t,
\hat\vb(t)) \Big)[\dot\Lambda_r^{t}]\\
&+\Big(\partial_\vb{\hat\tau}(\widehat\Lambda,
\Lambda_r^t,\hat\vb(t)+\wb)-\partial_\vb{\hat\tau}(\widehat\Lambda,\Lambda_r^t,\hat\vb(t))\Big)\cdot\dot{\vb}(t)
+{\widehat\db}_0(\widehat\Lambda,\Lambda_r^t)\cdot\wb=0.
\end{split}
\end{equation}

\medskip

\emph{Step 3. Energy Splitting.} As in \cite{PPG} for
instant-memory materials, we achieve the desired result for
materials with fading memory by showing that
$\partial_\Lambda\hat\tau$ does not depend on $\mathbf v$.

The argument used in \cite{PPG} relies on the existence, given an
arbitrary $\Omega$ in $L$, of a state process $\widehat\Gamma$
whose instantaneous value coincides with $\widehat\Lambda(t)$, and
whose rate $\dot\Gamma(t)$ equals $\Omega$; by replacing
$\widehat\Lambda$ with $\widehat\Gamma$, and by invoking the
arbitrariness of $\Omega$, one deduces that the term multiplying
$\dot\Lambda(t)$ in the first line of \eqref{eq:5} must vanish,
and then concludes that $\partial_\Lambda\hat\tau$ does not depend
on $\vb$. The argument we use in the present proof is similar, and
is based on a result due to Coleman and Mizel \cite[Remark 2]{CM}:
\emph{for every positive $\varepsilon$, there is a state process
$^\varepsilon\widehat\Gamma$ such that
\begin{align*}
 & ^\varepsilon\dot\Gamma(t)=\Omega,
\\
& |\widehat\Lambda(t)-\,
^\varepsilon\widehat\Gamma(t)|<\varepsilon,
\end{align*}
whose past history $^\varepsilon\Gamma^t_r(\cdot)$ satisfies
\begin{align*}
& \|\Lambda^t_r-\,^\varepsilon\Gamma^t_r(\cdot)\|<\varepsilon,
\\
&
\|\dot\Lambda^t_r-\,^\varepsilon\dot\Gamma^t_r(\cdot)\|<\varepsilon.
\end{align*}} \color{black}
Replacing $\widehat\Lambda$ with
$^\varepsilon\widehat\Gamma(\cdot)$ in \eqref{eq:5}, letting
$\varepsilon$ vanish, and using the smoothness of $\hat\tau$ and
$\widehat{\mathbf d}^{\rm in}$, we obtain
\begin{equation}\label{eq:6}
\begin{split}
&\Big(\partial_\Lambda{\hat\tau}(\widehat\Lambda(t),\Lambda_r^t,
\hat\vb(t)+\wb)-\partial_\Lambda{\hat\tau}(\widehat\Lambda(t),\Lambda_r^t,\hat\vb(t))\Big)\cdot\Omega+\dots=0
\end{split}
\end{equation}
(the dots stand for the remaining terms of \eqref{eq:5}), whence,
by the arbitrariness of $\Omega$,
\begin{equation}\label{eq:13}
\partial_\Lambda{\hat\tau}(\widehat\Lambda,\Lambda_r^t,
\hat\vb(t)+\wb)-\partial_\Lambda{\hat\tau}(\widehat\Lambda,\Lambda_r^t,\hat\vb(t))=\mathbf{0},
\quad \textrm{for all $\wb$ in $V$,}
\end{equation}
that is to say, as anticipated, $\partial_\Lambda\hat\tau$ does
not depend on the velocity. By \eqref{eq:13}, we conclude that the
constitutive mapping that delivers the total energy splits as
follows:
\begin{equation}\label{eq:6}
{\hat\tau}(\widehat\Lambda(t),\Lambda_r^t,\hat\vb(t))=\hat\epsilon(\widehat\Lambda(t),\Lambda_r^t)+\hat\kappa(\Lambda_r^t,\hat\vb(t)).
\end{equation}

\emph{Step 4. Representations of Kinetic Energy and Inertial
Force.} By \eqref{eq:6}, relation \eqref{eq:5} becomes:
 \begin{equation}\label{eq:11}
 \begin{split}
 &\Big(\delta_r\hat\kappa(\Lambda_r^t,\hat\vb+\wb)-\delta_r
 \hat\kappa(\Lambda_r^t,\hat\vb) \Big)[\dot\Lambda_r^t]
 +\Big(\partial_\vb\hat\kappa(\Lambda_r^t,\hat\vb+\wb)-\partial_\vb\hat\kappa(\Lambda_r^t,\hat\vb)
 \Big)\cdot\dot{\vb}+\\
 &+{\widehat\db}_0(\widehat\Lambda(t), \Lambda_r^t)\cdot\wb=0
 \end{split}
 \end{equation}
(the dependence of $\vb$ and $\dot\vb$ on $t$ has been left
tacit).

By assumption, both mappings $(\Lambda,\Lambda_r^t,\vb)\mapsto
\delta_r\hat\tau(\Lambda,\Lambda_r^t,\vb)[\,\cdot\,]$, and
$(\Lambda,\Lambda_r^t,\vb)\mapsto
\partial_{\vb}\hat\tau(\Lambda,\Lambda_r^t,\vb)$ are continuously
differentiable, hence\footnote{The identity \eqref{schw}
generalizes the theorem on the inversion of the order of partial
differentiation for functions of real variables \cite{Gr}.};
\begin{equation}\label{schw}
 \partial_{\mathbf v}(\delta_r\hat\tau(\Lambda,\Lambda_r^t,\vb)[\Phi_r])=\delta_r(\partial_{\vb}\hat\tau(\Lambda,\Lambda_r^t,\vb))[\Phi_r].
\end{equation}
On differentiating
 \eqref{eq:11} with respect to $\wb$ and using \eqref{schw} we obtain:
 \begin{equation}\label{eq:9}
 \delta_r\partial_{\vb}\hat\kappa(\Lambda_r^t,\hat\vb+\wb)[\dot\Lambda_r^t]+\partial_{\vb\vb}\hat\kappa(\Lambda_r^t,\hat\vb+\wb)\dot{\vb}+{\widehat\db}_0(\widehat\Lambda(t), \Lambda_r^t)=\0.
 \end{equation}
Upon choosing $\hat\vb=-\wb$ in \eqref{eq:9}, we have
 \begin{equation}\label{eq:10}
 \delta_r\partial_{\vb}\hat\kappa(\Lambda_r^t,\mathbf{0})[\dot\Lambda_r^t]+\partial_{\vb\vb}\hat\kappa(\Lambda_r^t,\mathbf{0})\dot{\vb}+{\widehat\db}_0(\widehat\Lambda(t), \Lambda_r^t)=\0.
 \end{equation}
 Finally, on subtracting \eqref{eq:10} from \eqref{eq:9} and selecting $\wb=\0$,
 we obtain:
\begin{equation}\label{final}
\Big(\delta_{r}\partial_{\vb}\hat\kappa(\Lambda_r^t,\hat\vb)-\delta_{r}\partial_{\vb}\hat\kappa(\Lambda_r^t,\0)\Big)[\dot\Lambda_r^{t}]
+\Big(\partial_{\vb\vb}\kappa(\Lambda_r^t,\hat\vb)-\partial_{\vb\vb}\kappa(\Lambda_r^t,\0)
\Big)\dot{\vb}=\0.
\end{equation}\color{black}
Since the choice of $\dot\vb$ is arbitrary, we conclude that
\begin{equation}\label{quadratic}
\hat\kappa(\Lambda_r^t,\vb)=\frac{1}{2}\hat\vb\cdot{\widehat\Mb}(\Lambda_r^t)\hat\vb,
\end{equation}
where the map ${\widehat\Mb}(\Lambda_r^t)$ takes its values in the
space of second-order symmetric tensors. By combining \eqref{eq:6}
and \eqref{quadratic}, we arrive at
\begin{equation}\label{reppre2}{
{\hat\tau}(\widehat\Lambda,\Lambda_r^t,\vb)=\hat\epsilon(\widehat\Lambda,\Lambda_r^t)+\frac{1}{2}\hat\vb\cdot{\widehat\Mb}(\Lambda_r^t)\hat\vb,}
\end{equation}
an additive energy splitting that generalizes the one in
\eqref{reppre}.

Substituting \eqref{quadratic} into \eqref{final}, we obtain
\begin{equation}\label{eq:12}
\delta_{r}{\widehat\Mb}(\Lambda_r^t)\,[\dot\Lambda_r^{t}]=0,
\end{equation}
and, by \eqref{eq:10},
\[
{\widehat\db}_0(\widehat\Lambda,\Lambda_r^t)=-{\widehat\Mb}(\Lambda_r^t)\dot{\vb}.
\]
From the linearity in the velocity of the power expenditure of
inertia force, we obtain the generalization of \eqref{finrep}
going alongside with \eqref{reppre2}:
\begin{equation}\label{finrep2}{
\widehat\db^{in}(\widehat\Lambda,\Lambda_r^t,\hat\vb)=\widehat\db_0(\widehat\Lambda,\Lambda_r^t)+\widehat\Db^{in}(\widehat\Lambda,\Lambda_r^t,\hat\vb)\hat\vb,
}
\end{equation}
where $\widehat\Db^{in}(\widehat\Lambda,\Lambda_r^t,\hat\vb)$ is a
skew-symmetric tensor. Finally, for
$\Mb(t)={\widehat\Mb}(\Lambda_r^t)$, the Chain-Rule Formula and
\eqref{eq:12} yield:
\begin{equation}\label{mdot}
 \dot \Mb(t)=0.
\end{equation}

\medskip

 We are now in position to wrap up our findings and state our
\medskip

 \noindent\textbf{Energy Splitting Theorem for
Materials with Fading Memory.} {\em Let the constitutive
dependence of energy and inertia be specified by
$\eqref{constitutive}_{1,2}$ and let the set of past histories be
endowed with the fading memory norm defined in \eqref{norm}.
Moreover, let the mappings
$(\widehat\Lambda,\Lambda^t_r,\vb)\mapsto\widehat\db^{\rm
in}(\widehat\Lambda,\Lambda^t_r,\vb)$ and
$(\widehat\Lambda,\Lambda^t_r,\vb)\mapsto\hat\tau(\widehat\Lambda,\Lambda^t_r,\vb)$
be, respectively, continuous and  twice-continuously
differentiable.

Assume that:\\
(i) the inertial power admit the representation \eqref{L1};\\
(ii) the total energy satisfy the invariance requirement
\eqref{I1}.

Then,  $\widehat\db^{\rm in}$ and $\hat\tau$ admit the representations
 \eqref{reppre2} and \eqref{finrep2}, parameterized by a symmetric-valued tensor mapping
 $\Lambda^t_r\mapsto\widehat\Mb(\Lambda^t_r)$, and a scalar mapping
$(\widehat\Lambda,\Lambda_r^t)\mapsto
\hat\epsilon(\widehat\Lambda,\Lambda_r^t)$. Moreover, the mass
tensor $\Mb(t)=\widehat\Mb(\Lambda^t_r)$ obeys the conservation
law \eqref{mdot}.}

\section{Materials with internal variables}
For materials with internal variables, the total energy, the
inertia force, and the work-effective part of the inertia force
are given by
\begin{equation}\label{constitutive1}
\begin{aligned}
\tilde\tau_\vb(t)&:=\hat\tau\big(\widehat\Lambda(t),\bbeta(t),\hat\vb(t)\big),\\
\widetilde\db_\vb^{\rm
in}(t)&:=\widehat\db^{in}\big(\widehat\Lambda(t),\bbeta(t),\hat\vb(t)\big),\\
\widetilde\db_0(t)&:=\widehat\db_0\big(\widehat\Lambda(t),\bbeta(t)\big)\color{blue},\color{black}
\end{aligned}
\end{equation}
where $\bbeta$ is a translationally-invariant vector of internal
variables, whose evolution is ruled by the ordinary differential
equation:
\begin{equation}\label{eq:evolution_internal}
\dot\bbeta(t)=\fb(\Lambda(t),\bbeta(t)).
\end{equation}
By \eqref{constitutive1} and
\eqref{eq:evolution_internal}, the invariance statement \eqref{timediff} leads to
 \begin{equation}\label{eq:1}
 \begin{split}
 &\Big(\partial_{\Lambda}{\hat\tau}(\widehat\Lambda(t),\bbeta(t),\hat\vb(t)+\wb)-\partial_{\Lambda}{\hat\tau}
 (\widehat\Lambda,\bbeta(t),\hat\vb(t))\Big)\cdot\dot{\Lambda}(t)\\
&+\Big(\partial_{\bbeta}\hat\tau(\widehat\Lambda(t),\bbeta(t),\hat\vb(t)+\wb)-\partial_{\bbeta}\hat\tau(\widehat\Lambda,
 \bbeta,\hat\vb) \Big)\cdot \fb(\Lambda(t),\bbeta(t))\\
 &+\Big(
 \partial_{\vb}\hat\tau(\widehat\Lambda(t),\bbeta(t),\hat\vb(t)+\wb)-\partial_{\vb}\hat\tau(\widehat\Lambda(t),\bbeta(t),\hat\vb(t))
 \Big)\cdot\dot{\vb}(t)\\
 &+\widehat\db_0(\widehat\Lambda(t), \bbeta)\cdot\wb=0.
 \end{split}
 \end{equation}
By a continuation argument borrowed from \cite{PPG}, the time
derivative $\dot\Lambda(t)$ appearing in the first line of
\eqref{eq:1} can be replaced with an arbitrary rate $\Omega$; the
resulting relation allows one to conclude that
 \[
 \partial_{\Lambda}{\hat\tau}(\widehat\Lambda(t),\bbeta(t),\hat\vb(t)+\wb)-\partial_{\Lambda}{\hat\tau}(\widehat\Lambda(t),\bbeta(t),\hat\vb(t))=\0
 \qquad \textrm{for all $\wb$ in $V$},
 \]
 whence the splitting:
 \[
 {\hat\tau}(\widehat\Lambda,\bbeta(t),\hat\vb(t))=\hat\epsilon(\widehat\Lambda(t),\bbeta(t))+\hat\kappa(\bbeta(t),\hat\vb(t)).
 \]
From this point on, the proof proceeds along the steps listed in
the previous section, with $\bbeta(t)$ in the place of
$\Lambda^t_r$. The conclusions  are, \emph{mutatis mutandis}, the
same as those stated in the Energy Splitting Theorem for materials
with fading memory: the total energy and the inertial force admit
the representations:
\[
    \hat\tau(\widehat\Lambda,\bbeta,\hat\vb)=\hat\epsilon(\widehat\Lambda,\bbeta)+{1\over 2}\,\hat\vb\cdot \widehat\Mb(\bbeta)\hat\vb\,,
\]
\[
    \widehat\db^{\rm in}(\widehat\Lambda,\bbeta,\hat\vb)=-\widehat\Mb(\bbeta)\dot \vb + \widehat\Db^{\rm in}(\widehat\Lambda,\bbeta,\hat\vb)\hat\vb,
\]
with $\widehat\Mb(\bbeta)$ symmetric and $\widehat\Db^{\rm
in}(\widehat\Lambda,\bbeta,\hat\vb)$ skew; the mass tensor
$\Mb(t)=\widehat\Mb(\bbeta(t))$ satisfies the conservation law
\eqref{mdot}.




\end{document}

%% file: caratteri.tex
%
%
%
%
\newcommand{\bydef}{\,\raise.050ex\hbox{\rm:}\kern-.025em\hbox{\rm=}\,}
\newcommand{\defby}{=\raise.075ex\hbox{\kern-.325em\hbox{\rm:}}\,}
\newcommand{\mtrp}  {{-\!\top}} 
\newcommand{\bdot}  {{\scriptscriptstyle\bullet}}
\def\qed{\relax\ifmmode\hskip2em \Box\else\unskip\nobreak\hskip1em $\Box$\fi}
%
%
\newcommand {\eps} {\varepsilon} 
\newcommand {\vp} {\varphi}      
\newcommand {\0} {\textbf{0}}    
\newcommand {\1} {\textbf{1}}    
%
\newcommand {\Ac}  {\mathcal{A}}
\newcommand {\Bc}  {\mathcal{B}}
\newcommand {\Cc}  {\mathcal{C}}
\newcommand {\Dc}  {\mathcal{D}}
\newcommand {\Ec}  {\mathcal{E}}
\newcommand {\Fc}  {\mathcal{F}}
\newcommand {\Gc}  {\mathcal{G}}
\newcommand {\Hc}  {\mathcal{H}}
\newcommand {\Kc}  {\mathcal{K}}
\newcommand {\Ic}  {\mathcal{I}}
\newcommand {\Jc}  {\mathcal{J}}
\newcommand {\Lc}  {\mathcal{L}}
\newcommand {\Mc}  {\mathcal{M}}
\newcommand {\Nc}  {\mathcal{N}}
\newcommand {\Oc}  {\mathcal{O}}
\newcommand {\Pc}  {\mathcal{P}}
\newcommand {\Rc}  {\mathcal{R}}
\newcommand {\Sc}  {\mathcal{S}}
\newcommand {\Tc}  {\mathcal{T}}
\newcommand {\Uc}  {\mathcal{U}}
\newcommand {\Vc}  {\mathcal{V}}
\newcommand {\Wc}  {\mathcal{W}}
\newcommand {\Zc}  {\mathcal{Z}}
%
%
\newcommand {\ab} {\mathbf{a}}
\newcommand {\bb} {\mathbf{b}}
\newcommand {\cb} {\mathbf{c}}
\newcommand {\db} {\mathbf{d}}
\newcommand {\eb} {\mathbf{e}}
\newcommand {\fb} {\mathbf{f}}
\newcommand {\gb} {\mathbf{g}}
\newcommand {\hb} {\mathbf{h}}
\newcommand {\ib} {\mathbf{i}}
\newcommand {\kb} {\mathbf{k}}
\newcommand {\lb} {\mathbf{l}}
\newcommand {\mb} {\mathbf{m}}
\newcommand {\nb} {\mathbf{n}}
\newcommand {\pb} {\mathbf{p}}
\newcommand {\qb} {\mathbf{q}}
\newcommand {\rb} {\mathbf{r}}
\renewcommand {\sb} {\mathbf{s}}
\newcommand {\tb} {\mathbf{t}}
\newcommand {\xb} {\mathbf{x}}
\newcommand {\ub} {\mathbf{u}}
\newcommand {\vb} {\mathbf{v}}
\newcommand {\wb} {\mathbf{w}}
\newcommand {\zb} {\mathbf{z}}
\newcommand {\Ab} {\mathbf{A}}
\newcommand {\Bb} {\mathbf{B}}
\newcommand {\Cb} {\mathbf{C}}
\newcommand {\Db} {\mathbf{D}}
\newcommand {\Eb} {\mathbf{E}}
\newcommand {\Fb} {\mathbf{F}}
\newcommand {\Gb} {\mathbf{G}}
\newcommand {\Hb} {\mathbf{H}}
\newcommand {\Kb} {\mathbf{K}}
\newcommand {\Jb} {\mathbf{J}}
\newcommand {\Ib} {\mathbf{I}}
\newcommand {\Lb} {\mathbf{L}}
\newcommand {\Mb} {\mathbf{M}}
\newcommand {\Nb} {\mathbf{N}}
\newcommand {\Ob} {\mathbf{O}}
\newcommand {\Pb} {\mathbf{P}}
\newcommand {\Qb} {\mathbf{Q}}
\newcommand {\Rb} {\mathbf{R}}
\newcommand {\Sb} {\mathbf{S}}
\newcommand {\Tb} {\mathbf{T}}
\newcommand {\Vb} {\mathbf{V}}
\newcommand {\Wb} {\mathbf{W}}
\newcommand {\Xb} {\mathbf{X}}
\newcommand {\Zb} {\mathbf{Z}}
%

%
%
\newcommand {\ax} {\mathrm{a}}
\newcommand {\bx} {\mathrm{b}}
\newcommand {\cx} {\mathrm{c}}
\newcommand {\dx} {\mathrm{d}}
\newcommand {\ex} {\mathrm{e}}
\newcommand {\fx} {\mathrm{f}}
\newcommand {\gx} {\mathrm{g}}
\newcommand {\hx} {\mathrm{h}}
\newcommand {\kx} {\mathrm{k}}
\newcommand {\lx} {\mathrm{l}}
\newcommand {\mx} {\mathrm{m}}
\newcommand {\nx} {\mathrm{n}}
\newcommand {\ox} {\mathrm{o}}
\newcommand {\px} {\mathrm{p}}
\newcommand {\qx} {\mathrm{q}}
\newcommand {\rx} {\mathrm{r}}
\newcommand {\sx} {\mathrm{s}}
\newcommand {\tx} {\mathrm{t}}
\newcommand {\xx} {\mathrm{x}}
\newcommand {\yx} {\mathrm{y}}
\newcommand {\wx} {\mathrm{w}}
\newcommand {\ux} {\mathrm{u}}
\newcommand {\vx} {\mathrm{v}}
\newcommand {\zx} {\mathrm{z}}
\newcommand {\Ax} {\mathrm{A}}
\newcommand {\Bx} {\mathrm{B}}
\newcommand {\Cx} {\mathrm{C}}
\newcommand {\Dx} {\mathrm{D}}
\newcommand {\Ex} {\mathrm{E}}
\newcommand {\Fx} {\mathrm{F}}
\newcommand {\Gx} {\mathrm{G}}
\newcommand {\Hx} {\mathrm{H}}
\newcommand {\Kx} {\mathrm{K}}
\newcommand {\Jx} {\mathrm{J}}
\newcommand {\Ix} {\mathrm{I}}
\newcommand {\Lx} {\mathrm{L}}
\newcommand {\Mx} {\mathrm{M}}
\newcommand {\Nx} {\mathrm{N}}
\newcommand {\Ox} {\mathrm{O}}
\newcommand {\Px} {\mathrm{P}}
\newcommand {\Qx} {\mathrm{Q}}
\newcommand {\Rx} {\mathrm{R}}
\newcommand {\Sx} {\mathrm{S}}
\newcommand {\Tx} {\mathrm{T}}
\newcommand {\Vx} {\mathrm{V}}
\newcommand {\Wx} {\mathrm{W}}
\newcommand {\Xx} {\mathrm{X}}
\newcommand {\Yx} {\mathrm{Y}}
\newcommand {\Zx} {\mathrm{Z}}
%

%
%
\newcommand {\Real} {\mathbb{R}}
\newcommand {\Aos} {\mbox{$\scriptstyle\mathbb{A}$}}
\newcommand {\Ao} {\mathbb{A}}
\newcommand {\Bo} {\mathbb{B}}
\newcommand {\Co} {\mathbb{C}}
\newcommand {\Cop} {\mbox{$\scriptstyle\mathbb{C}$}}
\newcommand {\Do} {\mathbb{D}}
\newcommand {\Fo} {\mathbb{F}}
\newcommand {\Go} {\mathbb{G}}
\newcommand {\Io} {\mathbb{I}}
\newcommand {\Mo} {\mathbb{M}}
\newcommand {\Ko} {\mathbb{K}}
\newcommand {\No} {\mathbb{N}}
\newcommand {\Ro} {\mathbb{R}}
\newcommand {\So} {\mathbb{S}}
\newcommand {\To} {\mathbb{T}}
\newcommand {\Vo} {\mathbb{V}}
\newcommand {\Zo} {\mathbb{Z}}
%
\newcommand {\Xes} {\mathscr{X}}
\newcommand {\ges} {\mathscr{g}}
\newcommand {\wes} {\mathscr{W}}
%
%
\newcommand {\aef} {\mathfrak{a}}
\newcommand {\fef} {\mathfrak{f}}
\newcommand {\gef} {\mathfrak{g}}
\newcommand {\hef} {\mathfrak{h}}
\newcommand {\mef} {\mathfrak{m}}
\newcommand {\nef} {\mathfrak{n}}
\newcommand {\kef} {\mathfrak{k}}
\newcommand {\wef} {\mathfrak{w}}
\newcommand {\sef} {\mathfrak{s}}
\newcommand {\zef} {\mathfrak{z}}
\newcommand {\Def} {\mathfrak{D}}
\newcommand {\Fef} {\mathfrak{F}}
\newcommand {\Mef} {\mathfrak{M}}
\newcommand {\Nef} {\mathfrak{N}}
\newcommand {\Ref} {\mathfrak{R}}
\newcommand {\Sef} {\mathfrak{S}}
\newcommand {\Xef} {\mathfrak{X}}
%
%
\newcommand {\att} {\mathtt{a}}
\newcommand {\btt} {\mathtt{b}}
\newcommand {\ctt} {\mathtt{c}}
\newcommand {\dtt} {\mathtt{d}}
\newcommand {\ftt} {\mathtt{f}}
\newcommand {\gtt} {\mathtt{g}}
\newcommand {\mtt} {\mathtt{m}}
\newcommand {\ntt} {\mathtt{n}}
\newcommand {\htt} {\mathtt{h}}
\newcommand {\ptt} {\mathtt{p}}
\newcommand {\qtt} {\mathtt{q}}
\newcommand {\rtt} {\mathtt{r}}
\newcommand {\stt} {\mathtt{s}}
\newcommand {\ttt} {\mathtt{t}}
\newcommand {\vtt} {\mathtt{v}}
\newcommand {\wtt} {\mathtt{w}}
\newcommand {\ztt} {\mathtt{z}}
\newcommand {\Ftt} {\mathtt{F}}
\newcommand {\Stt} {\mathtt{P}}
\newcommand {\Wtt} {\mathtt{W}}
%
%
%
%
\newcommand {\alfab}     {\mathbf{\alpha}}
\newcommand {\betab}     {\mathbf{\beta}}
\newcommand {\gammab}    {\mathbf{\gamma}}
\newcommand {\deltab}    {\mathbf{\delta}}
\newcommand {\epsilonb}  {\mathbf{\epsilon}}
\newcommand {\epsb}      {\mathbf{\varepsilon}}
\newcommand {\zetab}     {\mathbf{\zeta}}
\newcommand {\etab}      {\mathbf{\eta}}
\newcommand {\tetab}     {\mathbf{\teta}}
\newcommand {\vtetab}    {\mathbf{\vartheta}}
\newcommand {\iotab}     {\mathbf{\iota}}
\newcommand {\kappab}    {\mathbf{\kappa}}
\newcommand {\lambdab}   {\mathbf{\lambda}}
\newcommand {\mub}       {\mathbf{\mu}}
\font\mbo=cmmib10 scaled \magstephalf
\newcommand{\nub}   {\hbox{\mbo {\char 23}}}
\newcommand {\csib}      {\mathbf{\xi}}
\newcommand {\xib}      {\mathbf{\xi}}
\newcommand {\pib}       {\mathbf{\pi}}
\newcommand {\varrhob}   {\mathbf{\varrho}}
\newcommand {\sigmab}    {\hbox{\mbo {\char 27}}}
\newcommand{\taub}   {\hbox{\mbo {\char 28}}}
\newcommand {\upsilonb}  {\mathbf{\upsilon}}
\newcommand {\phib}      {\mathbf{\phi}}
\newcommand {\varphib}   {\mathbf{\varphi}}
\newcommand {\chib}      {\mathbf{\chi}}
\newcommand {\psib}       {\mathbf{\psi}}
\newcommand {\omegab}    {\mathbf{\omega}}
%
%
%
%
\newcommand {\Gammab}    {\mathbf{\Gamma}}
\newcommand {\Deltab}    {\mathbf{\Delta}}
\newcommand {\Tetab}     {\mathbf{\Theta}}
\newcommand {\Lambdab}   {\mathbf{\Lambda}}
\newcommand {\Csib}      {\mathbf{\Xi}}
\newcommand {\Pib}       {\mathbf{\Pi}}
\newcommand {\Sigmab}    {\mathbf{\Sigma}}
\newcommand {\Phib}      {\mathbf{\Phi}}
\newcommand {\Psib}      {\mathbf{\Psi}}
\newcommand {\Omegab}    {\mathbf{\Omega}}
%
%
\newcommand {\Lin} {\mathbb{L}\mathtt{in}}
\newcommand {\Sym} {\mathbb{S}\mathtt{ym}}
\newcommand {\Psym} {\mathbb{PS}\mathtt{ym}}
\newcommand {\Skw} {\mathbb{S}\mathtt{kw}}
\newcommand {\SO} {\mathcal{SO}}
\newcommand {\GL} {\mathcal{G}l}
\newcommand {\Rot} {\mathbb{R}\mathtt{ot}}
%
%
\newcommand {\tr}[1]{\mbox{tr}\, #1}
\newcommand {\psym} {\mbox{sym}}
\newcommand {\pskw} {\mbox{skw}}
\newcommand {\win}[2] {( #1 \cdot #2 )_\wedge }
\newcommand {\modulo}[1] {\left|#1\right|}
\newcommand {\sph} {\mbox{sph}}
\newcommand {\dev} {\mbox{dev}}
\newcommand {\sgn} {\mbox{sgn}}
\newcommand {\lin} {\mbox{Lin}}
%
%
\newcommand{\dvg} {\mathrm{div}\,}
\newcommand{\dvgt} {\mathrm{d{\widetilde i}v}\,}    
\newcommand{\grd} {\mathrm{grad}\,}    
\newcommand{\Grd} {\mathrm{Grad}\,}    
\newcommand{\dl}  {\delta}             
\def\gradtwo{\mathord{\nabla^{\scriptscriptstyle(2)}}}
\def\Div{\mathop{\hbox{Div}}}
\def\div{\mathop{\hbox{div}}}
\def\mis{\mathop{\hbox{mis}}}
\def\eps{\varepsilon}
%
%

\newcommand\ph{\varphi}
\newcommand{\adj} {\att\dtt}     

\newcommand{\va}{\mathbf{a}}
\newcommand{\vn}{\mathbf{\nu}}
\newcommand{\vt}{\mathbf{\tau}}
\newcommand{\dn}{\partial_{\mathbf{\nu}}}
\newcommand{\dt}{\partial_{\mathbf{\tau}}}
\newcommand{\ord}{\scriptscriptstyle}
\newcommand{\tD}{\mathbf{E}}  
\newcommand{\tS}{\mathbf{S}}  
\newcommand{\tE}{\mathbb{C}}
\newcommand{\tPf}{\mathbb{P}}
\newcommand{\tPr}{\mathbf{P}}
\newcommand{\tC}{\mathbf{C}}
\newcommand{\tP}{{\scriptstyle\mathbb{C}}}   
\newcommand{\tDl}{\mathbf{C}}      

\newcommand{\Cuno}{\mathbf{c}_1}
\newcommand{\Cdue}{\mathbf{c}_2}
\newcommand{\veralf}{\mathbf{c}_\alpha} 
\newcommand{\verbet}{\mathbf{c}_\beta} 
\newcommand{\vz}{\mathbf{z}}
\newcommand{\sym}{\mathop{\mathrm{sym}}}

\newcommand{\f}{f}
\newcommand{\g}{g}
\newcommand{\h}{h}
\newcommand{\w}{w}
\renewcommand{\l}{l}

\renewcommand{\r}{r}
\newcommand{\s}{s}

\newcommand{\autof}{\mathrm{\skew 0\overline{w}}}
\newcommand{\W}{W}

\newcommand{\df}{f'}
\newcommand{\dg}{g'}
\renewcommand{\dh}{h'}
\newcommand{\ddf}{f''}
\newcommand{\ddg}{g''}
\newcommand{\ddh}{h''}

\newcommand\modv[1]{|{#1}|}

\newcommand\arr[1]{\overrightarrow{#1}}
\newcommand{\cartref}{\{O;x_1,x_2,x_3\}}
\newcommand{\orthframe}{({\bf e}_1, {\bf e}_2, {\bf e}_3)}